\documentclass[sn-basic]{sn-jnl}

\jyear{2022}%

\usepackage{natbib}
\usepackage{graphicx}
\usepackage{mathrsfs}
\usepackage{color}
\usepackage{amssymb}
\usepackage{amsfonts}
\usepackage{amsmath}
\usepackage{bbm}
\usepackage{nameref}
\usepackage{stackengine}
\usepackage[pagewise]{lineno}
\usepackage{setspace}
\usepackage{multirow}
\usepackage{subcaption}
\usepackage{makecell}
\usepackage{placeins}

\definecolor{Gray}{gray}{0.9}

\allowdisplaybreaks
\usepackage{color, colortbl}
\usepackage{hyperref}

\begin{document}

\title[Remote work benefits and limitations]{\Large The limitations of remote work for reducing commuting-related carbon emissions}

\title[Caros, Guo and Zhao]{The benefits and limitations of remote work for reducing carbon emissions}

\author[1]{\fnm{Nicholas S.} \sur{Caros}}\email{caros@mit.edu}

\author[1]{\fnm{Xiaotong} \sur{Guo}}\email{xtguo@mit.edu}

\author[2]{\fnm{Jinhua} \sur{Zhao}}\email{jinhua@mit.edu}

\affil[1]{\small \orgdiv{Department of Civil and Environmental Engineering}, \orgname{Massachusetts Institute of Technology}, \orgaddress{\street{77 Massachusetts Avenue}, \city{Cambridge}, \postcode{02139}, \state{Massachusetts}, \country{United States of America}}}

\affil[2]{\small \orgdiv{Department of Urban Studies and Planning}, \orgname{Massachusetts Institute of Technology}, \orgaddress{\street{77 Massachusetts Avenue}, \city{Cambridge}, \postcode{02139}, \state{Massachusetts}, \country{United States of America}}}

\abstract{\normalsize 
Many studies of the effect of remote work on travel demand assume that remote work takes place entirely at home. 
Recent evidence, however, shows that in the United States, remote workers are choosing to spend approximately one third of their remote work hours outside of the home at cafe\'s, co-working spaces or the homes of friends and family. 
Commutes to these ``third places'' could offset much of the reduction in congestion and carbon emissions from commuting that could be expected from greater shares of remote work. 
To estimate the impact of third places on congestion and carbon emission from commuting, this study uses a national survey of thousands of remote workers and large-scale mobile trace data to predict current and future commuting patterns for the Chicago metropolitan area. 
The study reveals that ignoring third places leads to an underestimation of carbon emissions from commute-based travel demand by 470 gigatons per year, or 24\% of the total true emissions.
Moreover, if workers' latent desire for greater levels of remote work are realized in the future, the emissions benefits will be reduced further. 
The spatial analyses imply that there is a decrease in visits to the city center and outskirts, but an increase in visits to near suburban areas. 
Implications of these results for urban transportation and land use policy are discussed. 
}


\maketitle

\section*{Main}
\label{sec:intro}






Since the beginning of the COVID-19 pandemic, there has been considerable attention paid to the dramatic rise in working from home and the broader implications for society going forward.
However, the simple term ``working from home'' belies the fact that many workers have been spending their remote work hours in a wide range of places: coffee shops, libraries, co-working spaces, and friends' living rooms.
In this study, we demonstrate that the binary ``home-or-office as work locations'' paradigm fails to capture the true dynamics of remote work, and can lead to an overestimation of the benefits of remote work on two critical urban transportation indicators: total demand for travel and travel-related carbon emissions. 
Furthermore, we show how mobile phone data can be used to estimate commuting patterns for trips to non-home, non-work locations at a disaggregate level to facilitate long-term transportation planning.
Finally, we discuss the implications of these findings for urban land use and transportation policy in the U.S. context.

\begin{figure*}[ht!]
    \begin{center}
        \includegraphics[width=\textwidth]{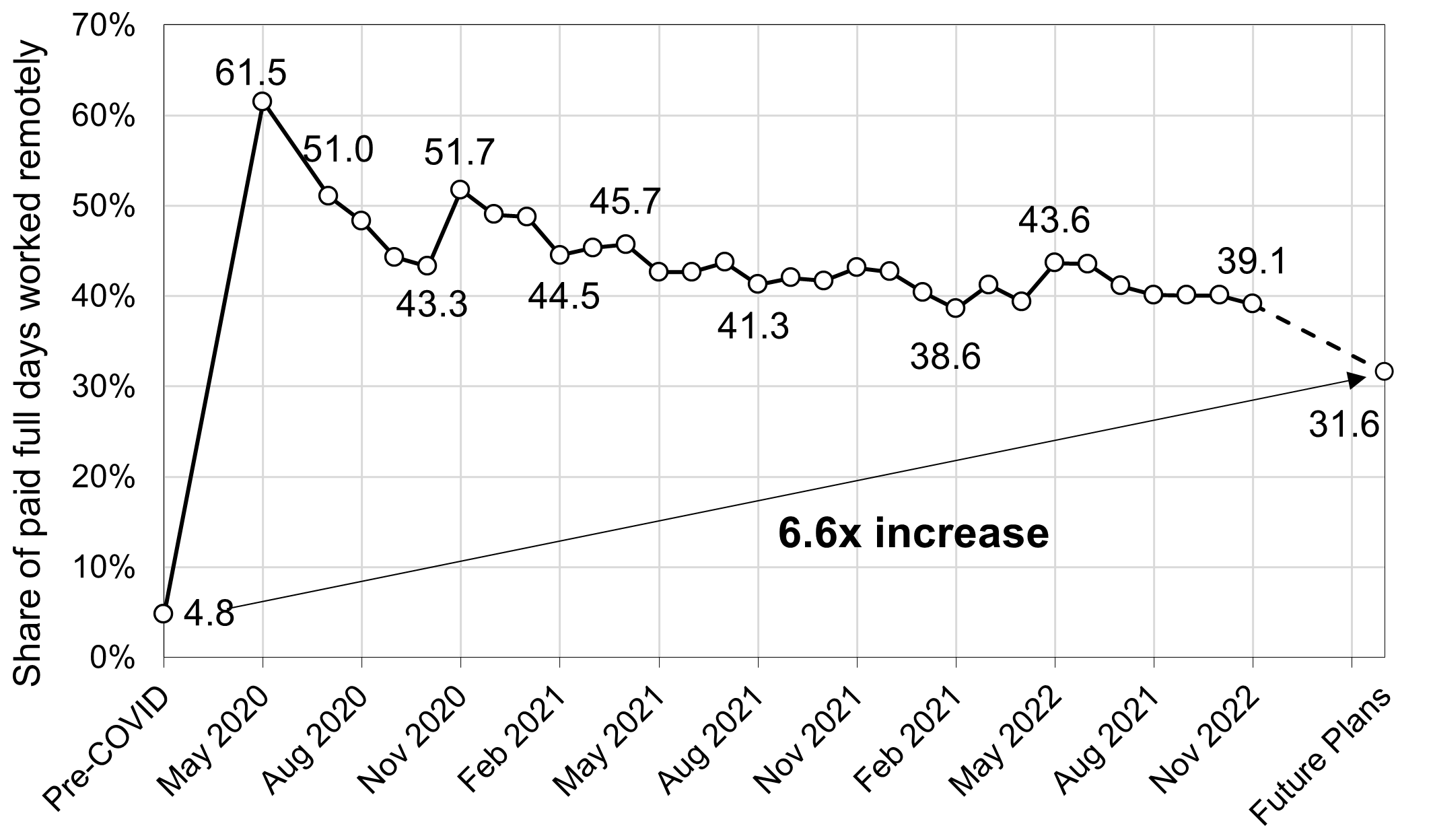}
        \caption{Flexible work trends before, during, and after the COVID-19 pandemic. Adapted from \cite{barrero2021working}.}
        \label{fig:remote_work}    
    \end{center}
\end{figure*}

From the beginning of the COVID-19 pandemic, there has been a strong research interest in the sudden increase in remote work adoption. 
In a working paper titled ``Why working from home will stick'', \citeauthor{barrero2021working}, show that remote work represented more than half of all worked hours in the United States during the height of the pandemic (see Figure~\ref{fig:remote_work}) \citep{barrero2021working}. 
The authors also find that remote work is expected to represent more than 31 percent of all worked hours after the pandemic subsides, a six-fold increase from 2018.

There has been a tremendous effort to understand the impact of remote work on travel demand since the outset of the COVID-19 pandemic.
Many studies have used survey instruments to elicit preferences for remote work across demographic groups during the pandemic's various stages \citep{beck2020easing, beck2020restrictions, shamshiripour2020covid, currie2021evidence, Salon_Conway_Capasso_2021, shibayama2021impact, thomas2021commuting}.
These surveys, while valuable for a range of research questions, only consider two possible working locations: home and a fixed workplace. 
Another set of articles used survey data to estimate statistical models for future commuting patterns, predicting a significant decline in overall commuting demand \citep{beck2020slowly,balbontin2021impact,hensher2021working, hensher2022impact}.
The models can be used for predicting travel modes and the number of commuting trips during non-remote working days, but each model assumes that remote work takes place entirely at home.

The narrow home vs. office framing of previous remote work studies can produce aggregate estimates of post-pandemic travel demand that ignore trips made to non-home remote work locations.
Even before the COVID-19 pandemic, it had been acknowledged that remote work was taking place in a variety of locations \citep{shearmur2021conceptualising}. 
In the post-COVID era, \cite{hensher2022impact} finds that many employers are beginning to embrace co-working spaces as an alternative to working from home, and \cite{beck2021might} points out that ``working close to home'' could be an appealing work modality.

Past literature, adapting a term from \cite{oldenburg1982third}, has referred to alternative remote work locations collectively as ``third places'' to differentiate them from the home and traditional workplace \citep{akhavan2021third, zenkteler2021role, mariotti2022were}; for consistency that terminology is also used here.
Understanding the use of third places is critical not only for travel behavior, but also for broad economic indicators such as employee satisfaction, firm productivity, and commercial real estate demand.

\begin{figure*}[ht!]
    \begin{center}
        \includegraphics[width=0.9\textwidth]{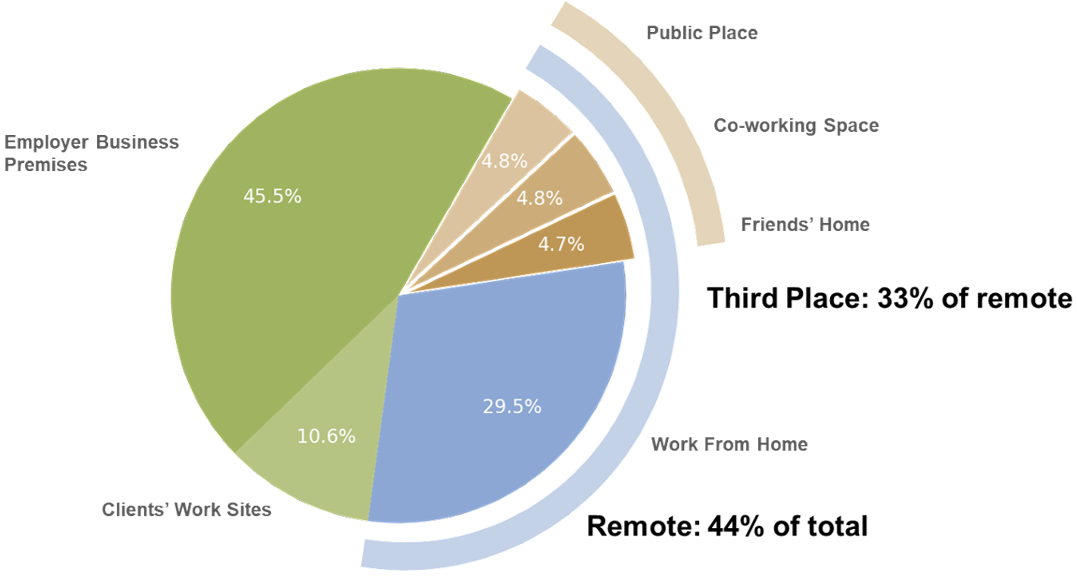}
        \caption{Distribution of worked hours by location type. Survey waves: November 2021 to March 2022. N = 21,136. }
        \label{fig:third_places}    
    \end{center}
\end{figure*}

To study the current and future use of third places, we have included several questions in recent waves of the Survey of Workplace Attitude and Arrangements (SWAA), a monthly survey of 5,000 working-age U.S. residents \citep{swaa}. 
We find that, after scaling the results to the demographics of the country, 14.3\% of total worked hours from November 2021 to March 2022 happened at a third place (see Figure~\ref{fig:third_places}).
This represents 32.6\% of all remote work hours in the United States.  
After weighting hours by income, we find that 36 percent of wages in the United States are earned in non-work, non-home locations.
The survey also shows that the distribution of employee preferences for third places is similar to their existing use of third places. 

Reported working hours at third places are relatively evenly split between the three categories included in the survey: public spaces (e.g. coffee shops or libraries), co-working spaces, and the home of a friend or family member.
Preferences for third places are not evenly distributed across the population, however. 
For example, the use of third places is more prevalent in urban areas than in suburban areas.
Similarly, the use of third places varies considerably by income group.
It is clear from the survey data that third places represent a significant proportion of remote work, with complex preferences that differ between demographic and employment groups. 
Quantifying the contribution of these factors as they relate to remote work preferences allows us to estimate the overall share of remote work as well as the preferences for third places for a given population.

We hypothesize that assuming all remote work takes place at home results in a significant underestimation of future travel demand and transportation-related carbon emissions. 
Our survey results demonstrate that third places are the chosen destination for a meaningful proportion of remote work commutes and those additional trips are currently being ignored.
Moreover, we conjecture that the false assumption leads to a skewed prediction of the spatial distribution of travel, as third place trips are typically shorter than a traditional commute and are more likely to take place within neighborhood centers.
This mischaracterization of travel demand could lead to insufficient sustainable transportation infrastructure, such as public transit or micromobility, to accommodate third place commutes. 
Ignoring remote work trips to third places is also predicted to overestimate the benefits of remote work with respect to reducing carbon emissions from commuting. 

In this study, we leverage a large, continuous nationwide survey to estimate preferences for remote work and third place visits for different demographic and geographic groups using Zero-One-Inflated Beta (ZOIB) regression and k-means clustering models. 
Then, we demonstrate how mobility trace data collected from a variety of sources can be used to estimate the characteristics of third place trips, including destination and distance.
Finally, we compute the carbon emissions related to traditional and third place commutes. 
This procedure allows us to quantify the effect of third places on aggregate travel demand, spatial demand patterns, and transportation-related carbon emissions across an urban area. 

The future is inherently uncertain, so we asked survey respondents three different questions about third place use.
The first question asks respondents to report their time spent working at a third place as a share of total work hours in order to estimate commuting patterns if there are no further changes in working arrangements. 
Then we ask about respondents' plans for working at third places in the medium-term future, assuming that the public health threat of the COVID-19 pandemic has subsided.
This provides the basis for a second scenario with each respondent's best guess for the future, including any future changes that their employer may be planning with regard to their working arrangements.
Finally, we ask about their desired time spent working at third places in the future, regardless of existing constraints, allowing us to develop a third hypothetical scenario in which workers are given total freedom over workplace choice. 
For each of the three scenarios, we compute the travel demand and carbon impacts with and without third place commutes and compare each against the pre-COVID baseline. 
This results in seven different possible commuting patterns that can be ranked against one another in terms of carbon emissions and total travel demand.

To the authors' knowledge, this study is the first of its kind to examine the impact of third places on post-pandemic travel demand, and to develop a method for forecasting the specific travel patterns resulting from an increase in remote work at third places by merging various data sources including surveys and mobile data.

\section*{Results}
\label{sec:results}

Results were generated using a systematic data-driven approach for estimating the new demand for travel under widespread remote work. 
First, the 2019 Chicago Household Travel Survey (CHTS) is used to examine commuting patterns before the pandemic. 
Then, to predict the individual levels of remote work and remote work location choices that affect commuting patterns, data from the longitudinal SWAA survey is incorporated. 
Finally, mobile phone records are combined with home location data to estimate the destinations of third place commuters. 

A four-step process was used to predict changes in travel patterns. 
The first is a ZOIB regression model estimated from the SWAA data to estimate individual shares of remote work. 
Next, a k-means clustering model is trained using the SWAA data to determine how remote work is divided between the home and different third place categories. 
Different questions from the SWAA survey provide an estimate for three scenarios: current (2022) levels of remote work, employees' desired levels of remote work, and employers' planned levels of remote work. 
Then, mobile phone record data are used to create a model that predicts the distribution of third place commute destinations based on the home census tract.
Finally, the results are aggregated for the entire urban area to generate multiple travel demand scenarios and calculate the carbon emissions associated with each scenario. 
To illustrate the importance of considering third places for remote work, we create a ``Home Only (HO)'' scenario where all remote work takes place at home and a more realistic ``Spectrum of Work Locations (SWL)'' scenario where some remote work occurs at third places.
\\

\begin{figure*}[h!]
    \centering
    \begin{subfigure}{\columnwidth}
        \centering
        \includegraphics[width=\textwidth]{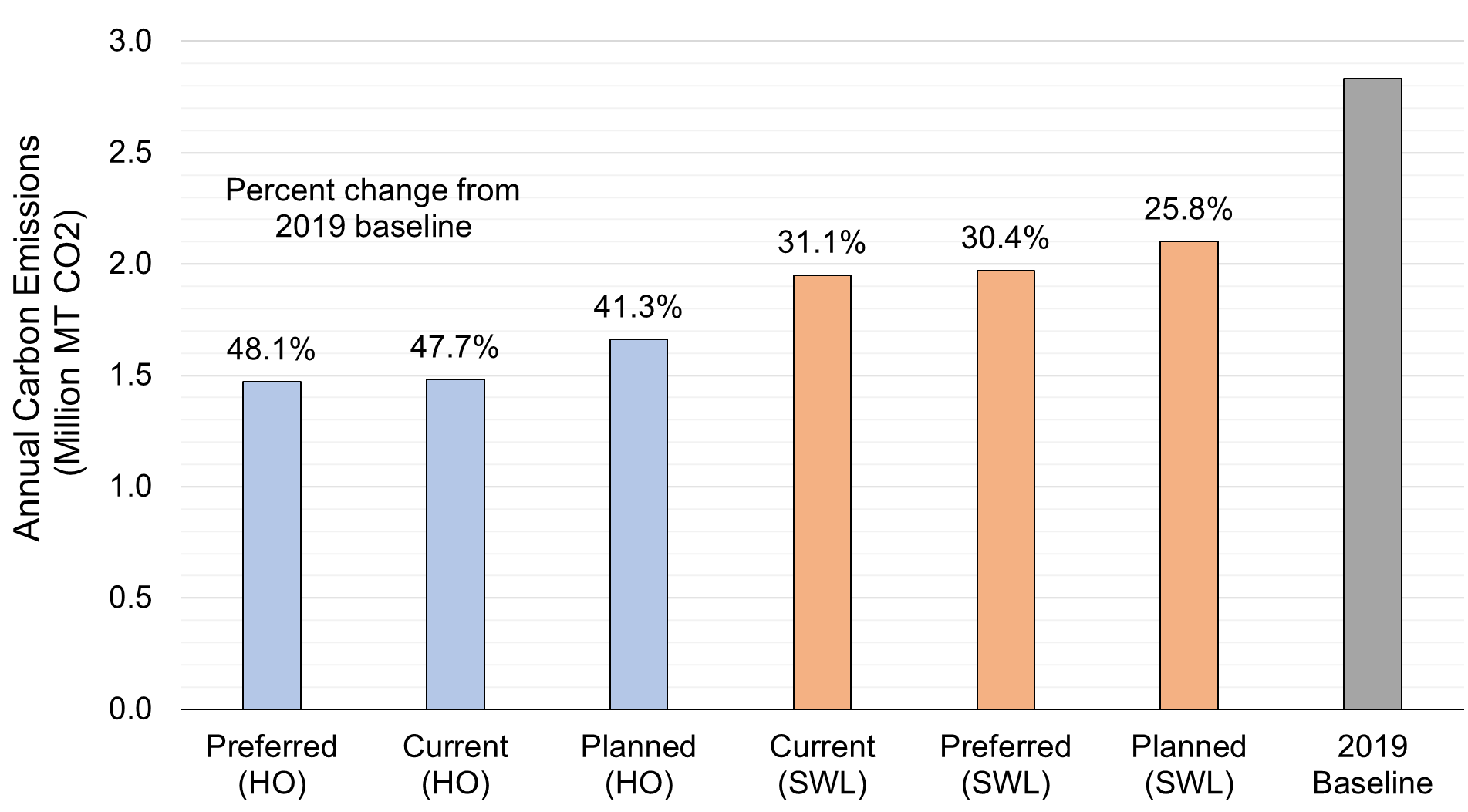}
    \end{subfigure}

    \vspace{0.15cm}

    \begin{subfigure}{\columnwidth}
            \centering
            \small
            \setlength{\tabcolsep}{4pt} 
            \resizebox{\textwidth}{!}{
            \begin{tabular}{|c|c|c|c|}
            \hline
            Scenario & \makecell{Annual Carbon Emission \\ (Million MT)} & Scenario & \makecell{Annual Carbon Emission \\ (Million MT)} \\
            \hline 
            Current (HO) & 1.48 ($-$47.7\%) & Current (SWL) & 1.95 ($-$31.1\%) \\
            \hline 
            Desired (HO) & 1.47 ($-$48.1\%) & Desired (SWL) & 1.97 ($-$30.3\%) \\
            \hline 
            Planned (HO) & 1.66 ($-$41.3\%) & Planned (SWL) & 2.10 ($-$25.8\%) \\
            \hline 
            \rowcolor{Gray} Pre-COVID Baseline & \multicolumn{3}{c|}{2.83} \\
            \hline 
            \end{tabular}}
    \end{subfigure}
    \caption{Carbon emissions for different commute-based travel demand scenarios}
    \label{fig:carbon_results}
\end{figure*}

\noindent \textbf{Carbon emissions from commuting.} 
As flexible work arrangements have increased from 4.8\% of worked hours pre-COVID to around 31.6\% in 2022, there has been a significant decrease in carbon emissions related to commuting as more people are working from home or at third places near their homes.
Table \ref{fig:carbon_results} presents the carbon emissions results for the six constructed scenarios and one pre-COVID baseline scenario.
The pre-COVID baseline was computed directly from CHTS data.
The constructed scenarios include three HO scenarios where people work only from home for flexible work and three SWL scenarios that consider commutes for remote work at third places. 
Within the HO or SWL scenarios, we determined travel patterns based on 1) current remote work rates and location choices, 2) employees' desired work from home rates and location choices, and 3) employers' planned work from home rates and location choices. 

Before discussing the results, it should be noted that this study is concerned only with estimating the carbon impact of changes to the length and frequency of \emph{commuting} trips to and from work as a result of the widespread increase in remote working.
It does not consider other important components of the overall influence of remote work on carbon emissions, such as non-work travel and building emissions.
The effect of remote work on the propensity for non-work travel has long been debated. 
Previous studies have found that under certain conditions, some remote workers conduct more non-work travel than those who work entirely in-person (e.g. \citet{deAbreuSilva_Melo_2018, zhu2018metropolitan,Su_McBride_Goulias_2021}), but other studies have found little-to-no effect under different conditions (e.g. \citet{choo2005does, Kim_Choo_Mokhtarian_2015, deAbreuSilva_Melo_2018b}). 
\citet{o2020does} provides an excellent summary of previous research on the various ``rebound effects'' of remote work, including changes to office and home energy consumption.
Their literature review shows that, like non-work travel, the impact of remote work on energy consumption for buildings is mixed and highly dependent on context and assumptions. 
While our study generates new insights into possible changes in commuting-related travel in the post-COVID era, it is but one piece of a holistic investigation into the overall carbon emissions impacts of widespread remote work. 

The model findings show that carbon emissions related to commuting have decreased by 31.1\% from the pre-COVID level. 
When third places are considered, people tend to work more at these locations and travel more while working remotely. 
For home-only scenarios, people generally prefer to work more flexibly from home rather than in the office. 
Employers are planning to have their employees work more frequently in the office, so scenarios based on employer plans produce fewer emissions relative to the employee preferences or current remote work scenarios. 
The results clearly demonstrate that it is important to consider third places when evaluating the impact of remote work on commuting, as not doing so can lead to an overestimation of the reduction in carbon emissions.
In the current scenario, which assumes remote work arrangements remain constant going forward, ignoring third places results in an underestimation of carbon emissions by 16.6\%. 

\begin{figure*}[h!]
    \begin{center}
        \includegraphics[width=1\textwidth]{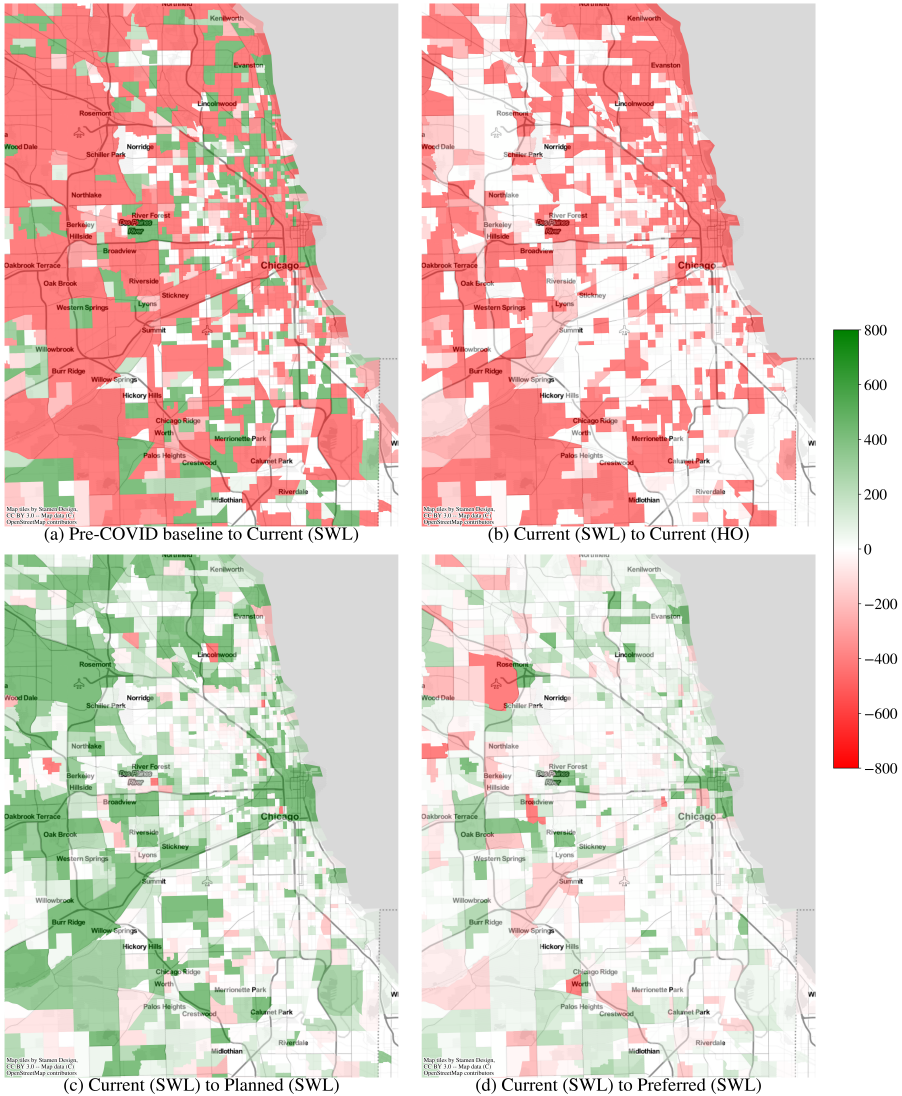}
        \caption{Changes in visits at census tract level between scenarios}
        \label{fig:spatial_maps}    
    \end{center}
\end{figure*}

\noindent \textbf{Spatial travel patterns.} By combining the actual work locations from the household travel survey with synthetic trips from mobile phone data, the model also estimates the disaggregate origin-destination patterns for each demand scenario. 
Figure \ref{fig:spatial_maps}(a) illustrates the change in the number of visits to each census tract from the pre-COVID baseline scenario to the current travel pattern (i.e. ``Current with SWL'' scenario). 
A ``donut effect'' can be observed, meaning that there is a decrease in visits to the city center and outskirts, but an increase in visits to near suburban areas.
These spatial patterns are reminiscent of the donut effect observed by \citet{ramani2021donut} with respect to housing prices after COVID-19. 
These results suggest that people are traveling more often to third places located in dense residential areas, rather than commuting to offices located in the commercial core.

Figure \ref{fig:spatial_maps}(b) illustrates the difference in visits between the current scenario with SWL and the current scenario with HO. 
It's clear that ignoring third places leads to the undercounting of many commuting trips, particularly in the densely-populated and amenity-rich northern part of Chicago.

Figure \ref{fig:spatial_maps}(c) shows the difference in visits between the current scenario and the employer-planned scenario that takes third places into account.
The employers in this scenario are planning to have their employees work in the office more frequently, resulting in more trips to the city center and outskirts and fewer trips to suburban areas. 
This is in contrast to the ``donut effect'' observed in Figure \ref{fig:spatial_maps}(a).

Figure \ref{fig:spatial_maps}(d) shows the difference in visits between the current scenario and the employee-desired scenario that takes third places into account. 
It's interesting to observe that people want to work more in the city center, outskirts, and certain suburban areas. 
These mixed results that in an ideal world, people would generally prefer to work slightly more at locations other than their homes, with some opting for the office and others choosing third places.

\FloatBarrier

\noindent \textbf{Remote work preferences.} While only one step in the aggregate travel demand process, the estimation of the ZOIB model provides several interesting insights into the dynamics of remote work. 
The estimated parameters can be used to determine the effect of the independent socioeconomic variables on the probability of choosing 0\% remote work, 100\% remote work, and the mean of the Beta distribution (denoted by $\mu$) if the proportion is neither 0\% nor 100\%. 
The results are shown for current, employee desired, and employer planned levels of remote work in Table~\ref{tab:zoib}. 
Note that the categorical variables ``sex'' and ``population density'' were set to Male and Urban for the reference group, respectively, and the median results are not scaled to the US population average. 

\begin{table}[!ht]
    \centering
    \small
    \caption{ZOIB regression results for the proportion of remote work}
    \label{tab:zoib}
    \setlength{\tabcolsep}{4pt} 
    \begin{tabular}{l |  c c c c c}

        \textbf{Variable} & \makecell{\textbf{P($x=0$)} \\ \textit{No} \\ \textit{Remote}} & \makecell{\textbf{P($x=1$)} \\ \textit{Fully} \\ \textit{Remote}} & \makecell{\textbf{P($0<x<1$)} \\ \textit{Hybrid} \\ \textit{Work}} & \makecell{\textbf{$\mathbf{\mu}$} \\ \textit{Hybrid} \\ \textit{Remote \%}}  & \makecell{\textbf{$\mathbb{E}[x]$} \\ \textit{Combined} \\ \textit{Effect}} \\ \hline
        \multicolumn{6}{l}{\textbf{Current Remote Work Share}} \\ \hline 
        \rowcolor{Gray} Median & 19.12\% & 17.09\% & 63.79\% & 55.38\% & 53.98\% \\ \hline
        Female & 4.27\% & 6.95 \% & -11.22\% & - & 0.23\% \\
        Suburban & 17.75\% & -2.46\% & -15.29\% & -2.28\% & -7.75\% \\
        Rural & 17.75\% & -2.46\% & -15.29\% & -2.21\% & -12.28\% \\
        Age (years) & 1.00\% & 0.13\% & -1.14\% & -0.12\% & -0.60\% \\
        Education (years) & -3.11\% & 0.11\% & 3.22\% & - & 1.82\% \\
        Income (\$10k) & -2.86\% & 0.06\% & 2.80\% & 0.16\% & 1.81\% \\ \hline
        
        \multicolumn{6}{l}{\textbf{Employee Desired Remote Work Share}} \\ \hline 
        \rowcolor{Gray} Median & 18.23\% & 20.30\% & 61.47\% & 60.34\% & 57.46\% \\ \hline
        Female & -0.61\% & 6.67 \% & -6.06\% & - & 2.72\% \\
        Suburban & - & - & - & -1.70\% & -0.44\% \\
        Rural & 8.87\% & -0.83\% & -8.04\% & -3.75\% & -6.74\% \\
        Age (years) & 0.57\% & 0.02\% & -0.60\% & -0.13\% & -0.40\% \\
        Education (years) & -3.37\% & 0.20\% & 3.17\% & - & 2.27\% \\
        Income (\$10k) & -1.59\% & -0.09\% & 1.68\% & 0.21\% & 1.06\% \\ \hline
        
        \multicolumn{6}{l}{\textbf{Employer Planned Remote Work Share}} \\ \hline 
        \rowcolor{Gray} Median & 28.41\% & 15.85\% & 55.74\% & 60.38\% & 49.02\% \\ \hline
        Female & - & - & - & - & - \\
        Suburban & - & - & - & -2.30\% & -0.54\% \\
        Rural & 10.74\% & -3.17\% & -7.57\% & -4.05\% & -8.73\% \\
        Age (years) & 0.59\% & 0.01\% & -0.60\% & -0.10\% & -0.40\% \\
        Education (years) & -2.65\% & 0.10\% & 2.76\% & - & 1.69\% \\
        Income (\$10k) & -1.89\% & 0.02\% & 1.86\% & 0.15\% & 1.27\% \\ \hline
        \multicolumn{6}{l}{Note: the - symbol represents a parameter that is not statistically significant} \\
        \multicolumn{6}{l}{at a 95\% confidence level.}
        
    \end{tabular}
\end{table}

For the current remote work model, the negative coefficients for the Education continuous variable with respect to ``No Remote'' and ``Fully Remote'' suggest that people with more education are more likely to work a hybrid work schedule. 
This model also indicates that women are more likely to be working either fully remotely or fully in person than men. 
Insignificant parameters in the employee preference model and employer plan model also have interesting implications.
The insignificance of the Surburban categorical variable implies that, unlike high and low density areas, living in a moderate density area does not have a statistical effect on preferences for different working arrangements. 
Similarly, in the employer plan model, the insignificance of the Female categorical variable suggests that gender does not play a statistically significant role in employers' plans for remote work. 

\begin{figure*}[h!]
    \begin{center}
        \includegraphics[width=\textwidth]{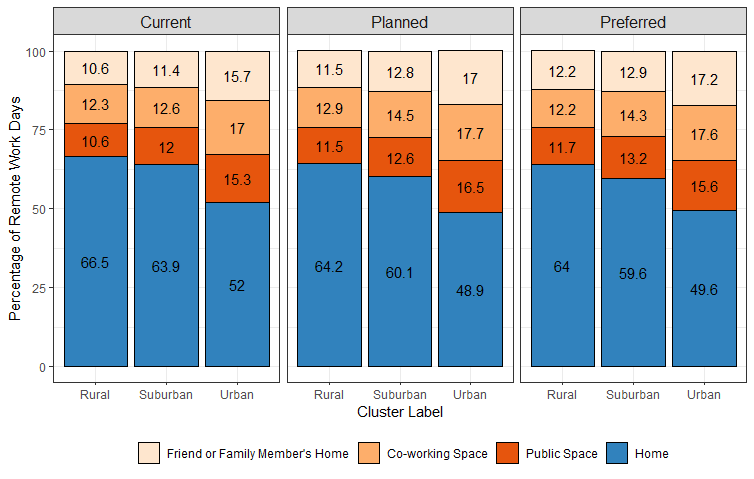}
        \caption{Current, employee desired, and employer planned remote work location distributions by cluster}
        \label{fig:clustering_results}    
    \end{center}
\end{figure*}

A k-means clustering approach is also trained on SWAA data to distribute this remote work share among different locations, including home, public space, friends' home, and co-working space. 
The remote work location choice probability distributions for each of the clusters are shown in Figure \ref{fig:clustering_results}.
The k-means clustering approach resulted in clusters that were largely differentiated by the home ZIP population density categories, with corresponding variations in the remaining socioeconomic and employment variables.
Remote workers in the ``urban'' cluster are much more likely to work at a third place than those in the ``suburban'' and ``rural'' clusters. 
These results seem sensible; low density land uses in rural and suburban areas make it more difficult for residents to access third places for remote work. 

\section*{Discussion}

The results of this study demonstrate that third place commuting trips are an important component of overall travel demand within a region, and should not be ignored. 
While remote work does reduce commuting overall compared to the pre-pandemic baseline, the impacts of remote work on commuting travel are somewhat dampened by trips to third places.
In addition, there is tension between employer plans and employee desires for remote work in the future.
If the tension is resolved in favor of the employers, then future travel demand is expected to be somewhat higher than in a compromise or employee preference-driven scenario. 
Third place commuting affects not only the aggregate level of travel and emissions, but also the spatial distribution of each measure.
These results have significant implications for transportation and land use planning going forward.

While there are some externalities as demonstrated by this study, the use of third places for remote work can have many positive effects on a community.
First, there is a travel cost for the commuter associated with visiting a third place, and many third places charge a fee (e.g. co-working space) or require a purchase (e.g. caf\'e) by the user. 
The fact that people choose to conduct remote work at third places despite these costs suggests that third places have some positive utility for remote workers relative to working from home. 
The revealed utility could be related to productivity, such as a less distracting environment compared to home or a stronger wi-fi connection. 
It might also be related to the opportunity to socialize or network with other remote workers, which can lead to spillover effects that boost the productivity of those involved.
Remote workers who choose to work at third places also support the third places and surrounding neighborhoods through economic activity. 

In a sense, the use of third places represents a compromise between working from home and working in a centralized employer-provided workplace. 
Remote workers benefit from a more social environment and avoid some of the negative aspects of working at home, while also limiting their own travel costs and the impact of their travel on others through shorter commutes.
The congestion and emission externalities of third place commutes can also be mitigated with intentional land use and transportation planning.
This study found that suburban and rural residents are less likely to use third places, and travel further when they do. 
Encouraging the development of third places outside of the city center would provide nearby options for remote workers in those areas, allowing them to reap the benefits of third places for remote work while reducing overall travel. 
The congestion externalities can be mitigated by providing sustainable transportation alternatives for third place trips to encourage the use of low-emissions modes. 
These alternatives could include better transit connections between residential areas and nearby town centers, or providing better micro-mobility, cycling, and walking infrastructure near third places. 
Third place commutes are also less likely peak hours compared to a traditional commute, so their impact on peak roadway and public transit congestion is of less concern. 

From a more general perspective, this paper proposes a rapid and inexpensive data-driven framework for revising travel demand estimates in the wake of sudden system-wide demand shocks. 
It leverages widely available, nationwide data sources that can be collected more quickly and with lower costs than a household travel survey.
We do not claim that the approach described herein is sufficient to replace household travel surveys altogether, as household surveys capture the granular data on individual trips needed to inform the data-driven approach.
However, it provides a useful first-order estimate of demand pattern changes that may occur in the years between household survey waves.

Many of the limitations of this study are related to data availability.
We used mobile records to infer destinations for third place trips, but future studies could collect these destinations directly from the survey respondents for an improved understanding of preferences for third places.
This area of research would also benefit from an exploration of alternative model structures for predicting remote work locations. 
The zero-one-inflated beta regression and clustering algorithms used in this study were selected for their accuracy, simplicity, and interpretability, but more complex models could be implemented in future research if suitable.
Finally, future work could explore policy prescriptions for reducing the impact of third places by optimizing zoning for third places near residential areas or developing operating strategies for public transit systems to serve third place commuters. 

\section*{Methods}
\subsection*{Data}
\label{sec:data}

There are three primary sources of data used in this analysis.
The first is the SWAA which is administered by a consortium of academic institutions \citep{swaa}.
The SWAA is the source of information for future remote work location preferences and includes demographic and employment data for each respondent.
The second is the My Daily Travel Survey conducted by the Chicago Metropolitan Agency for Planning (CMAP) between 2018 and 2019 \citep{cmap}. 
The CMAP survey includes detailed travel and personal information for over 12,000 households in the Chicago area and is available to the public.
This is the source of information for existing (pre-COVID) commuting patterns which are then modified based on the mode and work location changes predicted by the SWAA to produce an estimate of post-COVID commuting patterns.
Origin and destination locations for each trip are available at the census tract level.

The final source of information is SafeGraph, a data provider that aggregates anonymized location data from mobile applications in order to provide insights about travel and activity patterns.
Safegraph provides the locations of Points of Interest (POIs) and relative visitation frequencies for retail businesses \citep{safegraph}. 
SafeGraph information is used to determine the distribution of locations for trips by remote workers to third places such as coffee shops and co-working spaces.
Home locations for visitors in the SafeGraph dataset are available at the census block group level.

Note that the CMAP survey is the only data source that is specific to the Chicago area (the SWAA and SafeGraph are both national in scope).
Many state departments of transportation and Metropolitan Planning Organizations conduct similar surveys, so these results are largely generalizeable to other U.S. metropolitan areas subject to data availability.
The nationwide National Household Travel Survey could also be used to conduct a similar case study for the entire country, although doing so at the census tract or census block group level could present computational challenges.
Chicago was chosen to illustrate the methods presented in this study as it represents a very large urban area with high demographic and economic diversity. 

\subsection*{Third place impact}
\label{sec:methodology}

This study uses a four step procedure to evaluate the impact of third places on the demand for urban mobility at a disaggregate level. 
It begins with a baseline household travel survey, and seeks to update that survey to reflect changes in travel behavior. 
In this case, the primary changes in travel behavior are the substitution of traditional commuting trips with working at home or trips to third places. 
Specific data sources and methods are used to estimate how the travel behavior of each respondent in the baseline household travel survey change.
Then, the results are aggregated to provide an estimate of the overall impact of these travel behavior changes across the region. 

\begin{figure*}[ht!]
    \begin{center}
        \includegraphics[width=\textwidth]{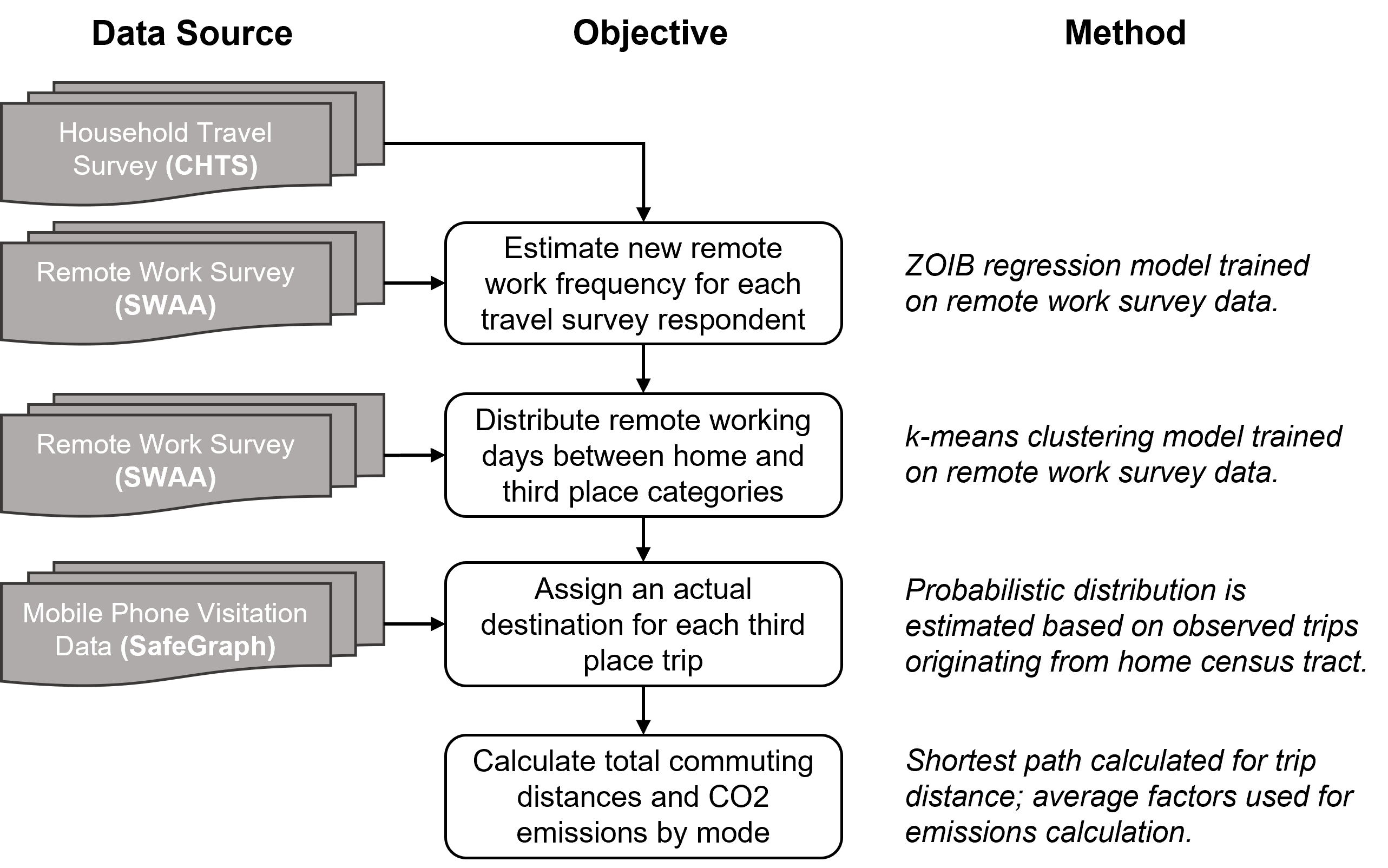}
        \caption{Flowchart demonstrating the emissions estimation process.}
        \label{fig:flowchart}    
    \end{center}
\end{figure*}

The overall procedure is summarized in Figure~\ref{fig:flowchart}.
Each of the steps are explained in detail in the subsections that follow. 

This procedure is similar to the canonical four-step model for travel demand forecasting \citep{de2011modelling}.
First, the number of total in-person and third place commuting trips for each individual is estimated, which is analogous to the ``trip generation'' step. 
Then, the third place trips are assigned to specific destinations, much like the ``trip distribution'' step of a traditional model.
Mode choice is extracted directly from the household survey data.
Route choice information is not available, so the shortest path with respect to travel time is assumed. 

\subsubsection*{Work location choice prediction (trip generation)}

To estimate how commuting patterns and carbon emissions could change as a result of working from third places, first we must predict the distribution of location choices for flexible work.
As discussed earlier, opportunities for remote work and preferences for third places are highly heterogeneous.
For that reason, a disaggregate approach is applied wherein work location choices are predicted for each individual using employment and demographic information.
A model is developed to predict, given a commuter with a specified set of demographic and employment variables, the fraction of pre-COVID commuting trips that fall into the following categories: A) eliminated due to working at home, B) have a modified destination due to working at a third place, and C) unchanged due to working at the employer's work site.
The trips within category B) are further distributed among the different types of third place: public space, friend's home and co-working space.

The scaled\footnote{SWAA responses are scaled to match the Current Population Survey (CPS) based on age, sex, education, and earnings. The detailed method can be found in \cite{barrero2021working}.} SWAA responses from November and December 2021 (N=7,950) are used as training data for the prediction model. 
The full list of SWAA employment and demographic variables used in the model are presented in Table~\ref{tab:inputs}.
Home ZIP Population density was split into three categories: Urban ($>3000$ residents per square mile), Suburban ($1000 - 3000$ residents per square mile) and Rural ($<1000$ residents per square mile).

\begin{table}[!ht]
    \centering
    \small
    \caption{Input variables for work location choice model}
    \label{tab:inputs}
    \begin{tabular}{l l}

        \textbf{Variable} & \textbf{Variable Type}  \\ \hline
        Sex & Categorical \\
        Age & Continuous \\
        Education & Continuous  \\
        Household Income & Continuous \\
        Home ZIP Population Density & Categorical \\

    \end{tabular}
\end{table}

\begin{figure*}[ht!]
    \begin{center}
        \includegraphics[width=\textwidth]{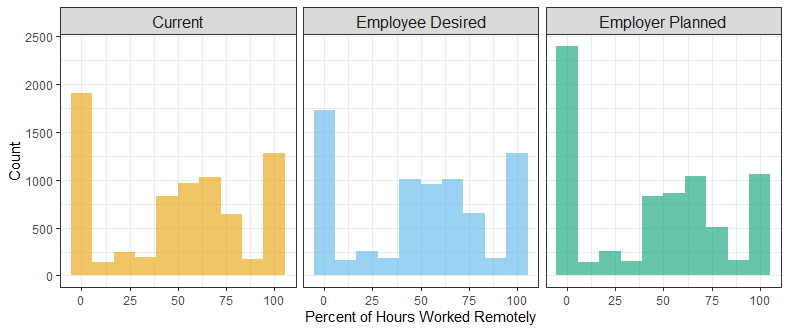}
        \caption{Histograms of current, employee desired, and employer planned remote work percentages}
        \label{fig:swaa_hist}    
    \end{center}
\end{figure*}

A ZOIB regression model was estimated using the SWAA data in order to predict the current, desired and planned percentages of remote work for each CHTS respondent.
ZOIB regression is a mixture model typically used to model proportion data where a qualitative difference between the populations with a 0\% response, a 100\% response and a response between 0\% and 100\% is expected. 
This conditions exists for the proportions of remote work.
The population with 0\% remote hours may represent a much different population than those working 1\% or more of their hours remotely. 
As an example, the 0\% population may work in a role where remote work is not possible (e.g. grocery store clerk, butcher, automotive mechanic), making it qualitatively different than the rest of the population.
Additionally, 100\% remote work enables a much different lifestyle than 90\% remote work by untethering the worker from the need to live near an office. 
As expected, the SWAA data is inflated at 0\% and 100\% of remote work hours as shown in Figure~\ref{fig:swaa_hist}.

The parameters of the three ZOIB model processes are estimated using Bayesian inference and presented in Table~\ref{tab:zoib}. 
After individual percentages of remote work are determined, a k-means clustering approach is employed to distribute this remote work share among different location types. 
In addition to the variables considered in the ZOIB model, the clustering model also incorporates categorical variables for work industry, occupation, and race. 
The Silhouette Coefficient was used to determine the most appropriate number of clusters for the dataset; the maximum value occurs when 3 clusters are used.

\subsubsection*{Assigning third place destinations (trip distribution)}

Creating new trips to third locations for future commuting patterns requires strong assumptions, but actual data was used wherever possible. 
There are three categories of third places in the SWAA questionnaire: public spaces, co-working spaces and the home of a friend or family member.
As the specific locations were not included in the survey, it is not possible to compute the distribution of travel distances for each of these location types directly.
An alternative means of estimating trip distances is therefore needed.

SafeGraph data uses mobile phone records to estimate the home locations of visitors to an extensive list of retail establishments. 
Establishment type is also included, so the distribution of visits from a given home location (at the census block group spatial resolution) to different establishment types can be determined.
This method is used to create a distribution of public space and co-working space visit probabilities and the associated travel distances for every home census tract in the Chicago Metropolitan Area. 
The expected value of public space and co-working space trip distance for each home census tract can then be estimated. 

Initial investigation found that third place trips were longer than expected due to noise in the SafeGraph data.
Unlike co-working spaces, visits to public spaces may be conducted for a variety of non-work reasons.
Visitors from distant suburbs may stop at a cafe\' as part of a shopping trip, for example.
Since the SafeGraph vistation data cannot be differentiated by trip purpose, a heuristic filter was applied to ensure that the estimated travel distances for remote work at public spaces reflect reasonable commuting behavior.
The filter removed any public space trips that exceed the length of the traditional commute to the employers' workplace by more than 1 kilometre, as it is unlikely remote workers would choose to travel further than their typical commute in order to work remotely from a public space. 

SafeGraph data does not contain information about visits to residential locations, so an alternative method is needed to estimate trip distances to the homes of friends and family members. 
Rather than SafeGraph data, the CMAP survey was used as it contains information related to the purpose of each trip.
Trips with the purpose of ``socializing with friends'' and ``socializing with relatives'' were used a proxy for visits to the homes of friends and family.
While socializing can take place in non-home locations, the inclusion of other trip purposes such as ``dining out'', ``shopping'', ``recreation'' and ``special event'' is assumed to reduce the number of non-home-based social events in the chosen trip categories. 
Aggregating the CMAP survey social trip distances for each home location census tract therefore provides a reasonable estimate of the distribution of travel distances for trips to friend's and family members' homes. 

The results for one-way trip distance by location type shown in Table~\ref{tab:trip_distances} demonstrate that third place commutes are typically much shorter than commutes to an employers' workplace.
The reduction in commuting distances for third places combined with the elimination of commutes altogether for at-home working days are the two drivers of the commuting-related carbon emissions reduction under widespread remote work.
If third place commuting distances were reduced further then future carbon emissions from commuting would be even lower.

\begin{table}[!ht]
    \centering
    \small
    \caption{Average one-way commuting trip distance by work location}
    \label{tab:trip_distances}
    \begin{tabular}{l | c }

         \textbf{Location type} & \textbf{Distance (km)} \\ \hline
         Employer workplace & 10.3 \\
         Public space & 3.5 \\
         Co-working space & 10.8 \\
         Friend or family member's home & 6.5 \\ \hline
        
    \end{tabular}
\end{table}

The average commuting distances to co-working spaces remains relatively high due to the concentration of available co-working space in the central business district of Chicago. 
The average commuting distance to public spaces, while the shortest of any location category, remains beyond a comfortable walking distance for most people.
Policies to reduce the average travel distances to third places would include zoning and incentives for locating new remote work-friendly public spaces and co-working spaces within residential areas.
Figure~\ref{fig:spatial_maps}(a) shows how census tracts to the west and south of downtown Chicago are not estimated to receive many third place commuting trips due to an lack of available destinations. 
Introducing new third places in such neighborhoods could be expected to have a disproportionately high impact on carbon emissions by offering a nearby destination for local third place commuters. 
Encouraging remote workers to use existing public spaces such as libraries and community centers would have a similar effect. 

\subsubsection*{Travel demand impacts}

Once the predicted change in commuting frequency and location distribution at the individual level is determined, the aggregate effects on travel demand can be calculated.
The 2018-2019 observed commuting distances are used as a baseline against which the predicted distances are compared.
Two scenario categories, HO and SWL, are compared to demonstrate the importance of including third place commutes in overall travel demand estimates. 
The overall change in aggregate travel distance and travel distance by mode is reported for both the HO and SWL scenarios. 
Furthermore, the overall change in trips by origin and destination are visualized by census tract for each scenario to identify spatial trends in third place commuting patterns. 

\subsubsection*{Carbon emissions}

There are two factors that contribute to the change in commute-related carbon emissions as a result of increased flexible work.
The first and most critical is the anticipated reduction in commuting distance that results from working at home and third places rather than a fixed employer-specific workplace. 
For the purpose of distance calculation it is assumed that trips to third places follow the shortest possible route.
This is a conservative assumption as a small number of travelers may choose deviate from the shortest path.
The results presented in this paper therefore represent an estimate of the lower bound of commuting-related carbon emissions when third place commutes are included.

The second factor that affects commuting-related carbon emissions is the change that arises from shifting from one commuting mode to another, as travel modes have significantly different emissions profiles. 
A targeted question was included in the January, February and March 2022 waves of the SWAA to determine whether remote workers use different travel modes depending on their choice of work location.
The survey found that individual remote workers almost always use the same travel mode whether commuting to a traditional workplace or one of the third place categories.
The aggregate mode share for traditional workplaces and third places are nearly identical.
As such, we make the modeling assumption that remote workers use the travel mode reported in the travel survey regardless of work location choice. 
The changes in carbon emissions from commuting are therefore influenced only by commuting distance. 

Using the difference in travel distance by mode from the previous section and multiplying by the average carbon emissions per unit distance by travel mode for the Chicago area, the total change in carbon emissions for both the HO and SWL scenarios is computed. The estimated $\text{CO}_2$ emissions per passenger mile for each travel mode provided by \citet{carbon_emission_param} is utilized to compute the total carbon emissions.


\bibliography{reference.bib}

\begin{thebibliography}{31}
\providecommand{\natexlab}[1]{#1}
\providecommand{\url}[1]{{#1}}
\providecommand{\urlprefix}{URL }
\providecommand{\doi}[1]{\url{https://doi.org/#1}}
\providecommand{\eprint}[2][]{\url{#2}}
 \bibcommenthead

\bibitem[{Akhavan(2021)}]{akhavan2021third}
Akhavan M (2021) {Third places for work: A multidisciplinary review of the
  literature on coworking spaces and maker spaces}. New Workplaces—Location
  Patterns, Urban Effects and Development Trajectories pp 13--32

\bibitem[{Balbontin et~al(2021)Balbontin, Hensher, Beck, Giesen, Basnak,
  Vallejo-Borda, and Venter}]{balbontin2021impact}
Balbontin C, Hensher DA, Beck MJ, et~al (2021) {Impact of COVID-19 on the
  number of days working from home and commuting travel: A cross-cultural
  comparison between Australia, South America and South Africa}. Journal of
  Transport Geography 96:103,188

\bibitem[{Barrero et~al(2021{\natexlab{a}})Barrero, Bloom, and
  Davis}]{barrero2021working}
Barrero JM, Bloom N, Davis SJ (2021{\natexlab{a}}) Why working from home will
  stick. Tech. rep., National Bureau of Economic Research

\bibitem[{Barrero et~al(2021{\natexlab{b}})Barrero, Bloom, Davis, and
  Zhao}]{swaa}
Barrero JM, Bloom N, Davis SJ, et~al (2021{\natexlab{b}}) {Survey of Working
  Arrangements and Attitudes}. \url{https://www.wfhresearch.com/data}, online;
  accessed on 2022-12-16

\bibitem[{Beck and Hensher(2020{\natexlab{a}})}]{beck2020easing}
Beck MJ, Hensher DA (2020{\natexlab{a}}) {Insights into the impact of COVID-19
  on household travel and activities in Australia -- The early days of easing
  restrictions}. Transport Policy 99:95--119

\bibitem[{Beck and Hensher(2020{\natexlab{b}})}]{beck2020restrictions}
Beck MJ, Hensher DA (2020{\natexlab{b}}) {Insights into the impact of COVID-19
  on household travel and activities in Australia -- The early days under
  restrictions}. Transport Policy 96:76--93

\bibitem[{Beck and Hensher(2021)}]{beck2021might}
Beck MJ, Hensher DA (2021) What might the changing incidence of working from
  home (wfh) tell us about future transport and land use agendas. Transport
  Reviews 41(3):257--261

\bibitem[{Beck et~al(2020)Beck, Hensher, and Wei}]{beck2020slowly}
Beck MJ, Hensher DA, Wei E (2020) {Slowly coming out of COVID-19 restrictions
  in Australia: Implications for working from home and commuting trips by car
  and public transport}. Journal of Transport Geography 88:102,846

\bibitem[{{Chicago Metropolitan Agency for Planning}(2020)}]{cmap}
{Chicago Metropolitan Agency for Planning} (2020) {My Daily Travel Survey,
  2018-2019}.
  \url{https://datahub.cmap.illinois.gov/dataset/mydailytravel-2018-2019-public},
  online; accessed on 2022-12-16

\bibitem[{Choo et~al(2005)Choo, Mokhtarian, and Salomon}]{choo2005does}
Choo S, Mokhtarian PL, Salomon I (2005) {Does telecommuting reduce
  vehicle-miles traveled? An aggregate time series analysis for the US}.
  Transportation 32(1):37--64

\bibitem[{Currie et~al(2021)Currie, Jain, and Aston}]{currie2021evidence}
Currie G, Jain T, Aston L (2021) {Evidence of a post-COVID change in travel
  behaviour -- Self-reported expectations of commuting in Melbourne}.
  Transportation Research Part A: Policy and Practice 153:218--234

\bibitem[{de~Dios~Ort{\'u}zar and Willumsen(2011)}]{de2011modelling}
de~Dios~Ort{\'u}zar J, Willumsen LG (2011) Modelling transport. John Wiley \&
  sons

\bibitem[{Hensher et~al(2021)Hensher, Beck, and Wei}]{hensher2021working}
Hensher DA, Beck MJ, Wei E (2021) {Working from home and its implications for
  strategic transport modelling based on the early days of the COVID-19
  pandemic}. Transportation Research Part A: Policy and Practice 148:64--78

\bibitem[{Hensher et~al(2022)Hensher, Balbontin, Beck, and
  Wei}]{hensher2022impact}
Hensher DA, Balbontin C, Beck MJ, et~al (2022) {The impact of working from home
  on modal commuting choice response during COVID-19: Implications for two
  metropolitan areas in Australia}. Transportation Research Part A: Policy and
  Practice 155:179--201

\bibitem[{Kim et~al(2015)Kim, Choo, and Mokhtarian}]{Kim_Choo_Mokhtarian_2015}
Kim SN, Choo S, Mokhtarian PL (2015) Home-based telecommuting and
  intra-household interactions in work and non-work travel: A seemingly
  unrelated censored regression approach. Transportation Research Part A:
  Policy and Practice 80:197–214

\bibitem[{Mariotti et~al(2022)Mariotti, Di~Matteo, and
  Rossi}]{mariotti2022were}
Mariotti I, Di~Matteo D, Rossi F (2022) {Who were the losers and winners during
  the Covid-19 pandemic? The rise of remote working in suburban areas}.
  Regional Studies, Regional Science 9(1):685--708

\bibitem[{O'Brien and Aliabadi(2020)}]{o2020does}
O'Brien W, Aliabadi FY (2020) Does telecommuting save energy? a critical review
  of quantitative studies and their research methods. Energy and buildings
  225:110,298

\bibitem[{Oldenburg and Brissett(1982)}]{oldenburg1982third}
Oldenburg R, Brissett D (1982) The third place. Qualitative Sociology
  5(4):265--284

\bibitem[{Ramani and Bloom(2021)}]{ramani2021donut}
Ramani A, Bloom N (2021) The donut effect of covid-19 on cities. Tech. rep.,
  National Bureau of Economic Research

\bibitem[{SafeGraph(2021)}]{safegraph}
SafeGraph (2021) {SafeGraph Places Data}. \url{https://www.safegraph.com/},
  online; accessed on 2022-12-16

\bibitem[{Salon et~al(2021)Salon, Conway, Capasso~da Silva, Chauhan, Derrible,
  Mohammadian, Khoeini, Parker, Mirtich, Shamshiripour, Rahimi, and
  Pendyala}]{Salon_Conway_Capasso_2021}
Salon D, Conway MW, Capasso~da Silva D, et~al (2021) {The potential stickiness
  of pandemic-induced behavior changes in the United States}. Proceedings of
  the National Academy of Sciences 118(27):e2106499,118

\bibitem[{Shamshiripour et~al(2020)Shamshiripour, Rahimi, Shabanpour, and
  Mohammadian}]{shamshiripour2020covid}
Shamshiripour A, Rahimi E, Shabanpour R, et~al (2020) {How is COVID-19
  reshaping activity-travel behavior? Evidence from a comprehensive survey in
  Chicago}. Transportation Research Interdisciplinary Perspectives 7:100,216

\bibitem[{Shearmur(2021)}]{shearmur2021conceptualising}
Shearmur R (2021) Conceptualising and measuring the location of work: Work
  location as a probability space. Urban Studies 58(11):2188--2206

\bibitem[{Shibayama et~al(2021)Shibayama, Sandholzer, Laa, and
  Brezina}]{shibayama2021impact}
Shibayama T, Sandholzer F, Laa B, et~al (2021) {Impact of COVID-19 lockdown on
  commuting: A multi-country perspective}. European Journal of Transport and
  Infrastructure Research 21(1):70--93

\bibitem[{de~Abreu~e Silva and
  Melo(2018{\natexlab{a}})}]{deAbreuSilva_Melo_2018b}
de~Abreu~e Silva J, Melo PC (2018{\natexlab{a}}) {Does home-based telework
  reduce household total travel? A path analysis using single and two worker
  British households}. Journal of Transport Geography 73:148–162

\bibitem[{de~Abreu~e Silva and
  Melo(2018{\natexlab{b}})}]{deAbreuSilva_Melo_2018}
de~Abreu~e Silva J, Melo PC (2018{\natexlab{b}}) Home telework, travel
  behavior, and land-use patterns: A path analysis of british single-worker
  households. Journal of Transport and Land Use 11(1):419–441

\bibitem[{Su et~al(2021)Su, McBride, and Goulias}]{Su_McBride_Goulias_2021}
Su R, McBride EC, Goulias KG (2021) Unveiling daily activity pattern
  differences between telecommuters and commuters using human mobility motifs
  and sequence analysis. Transportation Research Part A: Policy and Practice
  147:106–132

\bibitem[{Thomas et~al(2021)Thomas, Charlton, Lewis, and
  Nandavar}]{thomas2021commuting}
Thomas FM, Charlton SG, Lewis I, et~al (2021) {Commuting before and after
  COVID-19}. Transportation Research Interdisciplinary Perspectives 11:100,423

\bibitem[{{U.S. Department of Transportation Federal Transit
  Administration}(2010)}]{carbon_emission_param}
{U.S. Department of Transportation Federal Transit Administration} (2010)
  {Public Transportation’s Role in Responding to Climate Change}

\bibitem[{Zenkteler et~al(2021)Zenkteler, Foth, and Hearn}]{zenkteler2021role}
Zenkteler M, Foth M, Hearn G (2021) {The role of residential suburbs in the
  knowledge economy: Insights from a design charrette into nomadic and remote
  work practices}. Journal of Urban Design 26(4):422--440

\bibitem[{Zhu et~al(2018)Zhu, Wang, Jiang, and Zhou}]{zhu2018metropolitan}
Zhu P, Wang L, Jiang Y, et~al (2018) Metropolitan size and the impacts of
  telecommuting on personal travel. Transportation 45(2):385--414

\end{thebibliography}

\end{document}